**Title:** Architectured Chain Lattices with Tailorable Energy Absorption in Tension

**Authors:** Spencer V Taylor[1], Abdel R Moustafa[2], Zachary C Cordero[1]

**Affiliations:**

1. Aeronautics and Astronautics, MIT, Cambridge, MA
2. Materials Science and NanoEngineering, Rice University, Houston, TX

**Abstract:** This paper introduces the chain lattice, a hierarchical truss structure comprising two interpenetrating lattices. One lattice toughens the material and prevents catastrophic localized failure while the other lattice serves as a porous matrix that densifies to absorb energy during tensile loading. Chain lattices are amenable to additive manufacturing and can transform 3D-printable materials that are normally brittle and flaw-sensitive into damage-tolerant materials. Calculations predict ceramic chain lattices can have a specific energy absorption several orders of magnitude greater than that of their fully dense counterparts.

**Corresponding Author:** Zachary Cordero; zcordero@mit.edu; 617-253-8821

Energy absorption is the ability of a material to dissipate mechanical energy through inelastic deformation. In brittle materials, failure in tension occurs prior to appreciable inelastic work owing to their flaw sensitivity and low toughness. One way to improve the toughness of brittle materials is by developing composites. Fiber-reinforced brittle-matrix composites, for example, use fibers to bridge matrix cracks and distribute stress around strain-concentrating features [1,2]. In particular, woven and braided fiber-reinforced composites undergo local hardening events in which components of the reinforcement displace, then lock up to resist further local displacement. This localized hardening forces displacement to continue elsewhere in the composite, delocalizing damage and preventing catastrophic failure [3–5].

The hardening and damage delocalization behaviors of woven and braided brittle-matrix composites are exemplified in the chain composite, developed by Cox et al. [3,4,6–8]. Chain composites consist of parallel steel chains embedded in a brittle matrix. Under a tensile load, adjacent chain links, initially relaxed, slide towards one another as the chain is drawn tight, and the adjacent chain crowns compress trapped filler material upon displacement. Once the inner radii of the chains come into contact and experience lockup, further tensile deformation of the chain composite can only take place when filler in another location begins to compress. Finally, individual chain links and the entire composite begin to fail after all chain links have experienced lockup. Specific energy absorption for chain composites of steel chains and brittle polycarbonate filler measured up to 14 J/g, exceeding that of fiber-reinforced ceramic matrix composites (CMCs) by more than an order of magnitude and roughly equaling that of monolithic ductile alloys [7].

Brittle-matrix composites such as chain composites formed via conventional processing techniques often contain structural heterogeneities which limit their performance relative to theoretical upper bounds. Additive manufacturing (AM) offers a potential solution to this problem, since it can be used to fabricate complex structures with well-controlled geometries that impart specific, desirable failure behaviors. For example, AM has been used to precisely arrange the spatial distribution of constituents in interpenetrating phase composites (IPCs) [9,10], which enables IPCs to show tailored crack bridging [11] and damage delocalization [12,13]. This ability to manipulate failure behavior by patterning constituents can be extended to brittle materials via AM, motivating the work at present.

Here, we leverage the net-shaping capabilities of AM to develop a 3D-printable interpenetrating lattice structure called the chain lattice, depicted in **Figure 1**. The chain lattice is a composite of two lattices: the chain reinforcement and the filler lattice, where the chain reinforcement toughens the material and prevents catastrophic failure, while the filler lattice serves as a porous matrix that densifies to absorb energy during tensile loading. The chain reinforcement consists of links which displace towards one another under an applied tensile load, similar to the metal chain links of the chain composite. The filler lattice, printed simultaneously with the same material, fills the space between parallel chains and between the crowns of adjacent chain links, experiencing compression as the links displace. Thus, by patterning porosity within a brittle material, we can achieve chain lattices that exhibit localized hardening, tensile lockup, and delocalized damage.

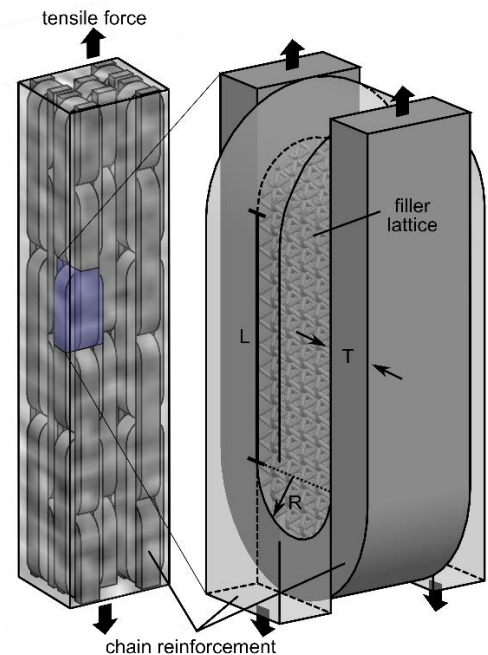

**Figure 1:** Chain lattice comprising two interpenetrating lattices of the same base material. Key design features (thickness $T$, length $L$, crown radius $R$) are labeled in the magnified view of the half-unit cell.

To design a chain lattice that exhibits delocalized damage, lockup must raise the strength of locally failed material such that failure occurs elsewhere [8]. This criterion dictates that the tensile load of the chain reinforcement exceeds the load required to compress the filler lattice, which can be expressed as

$$\sigma_r A_r \geq \sigma_f A_f, \tag{1}$$

where $\sigma_r$ is the tensile strength of the chain reinforcement, $\sigma_f$ is the compressive strength of the filler lattice, $A_r$ is the minimum area fraction of the chain reinforcement on all planes perpendicular to the tensile loading axis, and $A_f$ is the maximum area fraction of the filler lattice on all planes perpendicular to the tensile loading axis. Note that **Equation 1** does not account for effects due to interfacial bonding and shear, but only considers tension in the chain links due to compression of the filler lattice.

To satisfy this delocalized damage criterion for a given chain link geometry, we must accordingly tailor the strength of the filler lattice. Gibson and Ashby have described models to predict lattice compressive strength $\bar{\sigma}$ based on the degree of strut connectivity [14], with compressive strength given by

$$\bar{\sigma} \approx C_1 \sigma_o \bar{\rho}^n, \tag{2}$$

where $C_1$ is a proportionality constant, $\sigma_o$ is a material property, $\bar{\rho}$ is relative density of the lattice, and $n$ is a connectivity-dependent constant. For bend-dominated lattices (where strut connectivity is less than 12), $C_1 \approx 0.2$, $\sigma_o$ is the modulus of rupture of a strut, and $n \approx 1.5$; for stretch-dominated lattices (where strut connectivity is greater than 12), $C_1 \approx \frac{1}{3}$, $\sigma_o$ is the compressive strength of the fully dense material, and $n \approx 1$ [14,15].

To demonstrate the concept of a chain lattice, we printed a poly(methyl methacrylate) chain lattice using a Formlabs Form 2 stereolithography printer with their Clear Resin. We assessed the stress-strain behavior of the fully dense material, a body-centered cubic (BCC) lattice, and a chain lattice, all printed from the same material, through tension testing at room temperature using an electromechanical load frame with an initial strain rate of roughly $10^{-3}$ s$^{-1}$. The tension test specimens had a gauge length and gauge diameter of 5.2 cm and 1.3 cm, respectively. The stress-strain data is shown in **Figure 2a**. The fully dense material failed at a peak stress of 36 MPa due to the formation of a single crack perpendicular to the tensile loading axis, indicating its brittle nature. Based on the peak stress of the monolithic material, the filler lattice was designed as BCC with a relative density of 12%, which has bend-dominated behavior with compressive strength predicted as 0.3 MPa using **Equation 2**. In tension, the filler lattice failed sequentially, with a tensile strength on the order of 0.1 MPa. The steps in the stress-strain curve correspond to strut failure. Slight strain hardening occurred as struts rotated to align with the tensile axis.

Given the strengths of the fully dense and the filler lattice materials, we designed the chain lattice according to the delocalized damage constraint given in **Equation 1**. A sketch of the chain lattice tensile specimen is shown in **Figure 2b**, with a half chain link shown in **Figure 2c** and a unit cell of the filler lattice shown in **Figure 2d**. The maximum allowable ratio $\frac{A_f}{A_r}$ was 67, so we designed the chain lattice with $\frac{A_f}{A_r} = 0.74$, well within the bounds to produce delocalized damage. The chain link dimensions were $R = 5$ mm, $T = 5$ mm, and $L = 15$ mm; the filler lattice unit cell length was 2.5 mm, and the strut width was 0.35 mm; and the overall relative density of the chain lattice was 60%

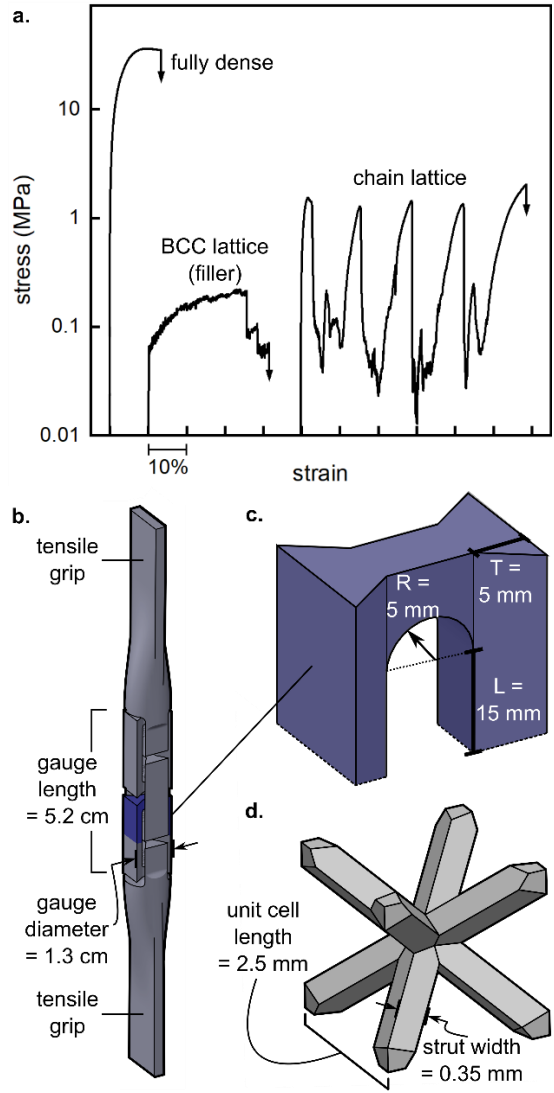

**Figure 2:** (a) Logarithmic tensile stress-strain data for 3D-printed poly(methyl methacrylate) in different forms – fully dense, a 12% dense BCC lattice, and a chain lattice with a 12% dense BCC filler lattice. (b) Chain lattice tensile specimen with gauge length = 5.2 cm and gauge diameter = 1.3 cm. (c) Half chain link of the chain lattice, where the dimensions were as follows: $R = 5$ mm, $T = 5$ mm, and $L = 15$ mm. (d) BCC filler lattice unit cell with relative density = 12%, unit cell length = 2.5 mm, and strut width = 0.35 mm.

In the stress-strain data of the chain lattice, the first peak of roughly 1 MPa corresponds to compressive yielding of the filler lattice. The resulting drop to roughly 0.1 MPa represents the crush-up stress, or the stress required to compress the fractured lattice until lockup between the two surrounding chain links occurred at roughly 10% strain. This process occurred a total of four times, corresponding to the four filler lattice domains in the specimen and embodied by the four peaks in the stress-strain data. After all the chains experienced lockup, failure occurred when one chain link failed in the junction between the leg and the chain crown. This lockup process notably led to a roughly 60% strain to failure in the chain lattice as compared to 12% and 32% seen in the monolithic material and the BCC lattice, respectively. The measured chain lattice strain to failure is greater than the predicted value of 40% obtained using Gibson and Ashby's approximation for lattice densification strain, given as [14]

$$\varepsilon \approx 1 - 1.4\bar{\rho}. \tag{3}$$

Our measurement likely exceeded the predicted value because fragmented material from the filler lattice was ejected from within the chain crowns during the test, allowing further displacement before the filler lattice densified.

While these results demonstrate the feasibility of chain lattices using an easy to process model system, their benefits become even more apparent in high-strength ideally brittle materials such as ceramics. In designing a ceramic chain lattice, the damage delocalization criterion must be modified to account for size effects on tensile strength due to Weibull statistics [16]. This effect stems from the probability of finding a flaw of critical size in a given volume of brittle material; critical flaws are more likely to be present in specimens with a larger volume for a given stress state, and the tensile strength of a ceramic material decreases with volume according to the relationship

$$\sigma_1 = \sigma_T \left(\frac{V_o}{V_1}\right)^{\frac{1}{m}}, \tag{4}$$

where $\sigma_T$ is the tensile strength of a representative volume $V_o$, $\sigma_1$ is the tensile strength of a specimen with volume $V_1$, and $m$ is the Weibull modulus, a material property that characterizes the volume effect on strength knockdown. Thus, with a constant chain lattice cross-sectional area, the tensile strength of the chain reinforcement decreases with increasing leg length $L$ (cf. **Figure 1**), and the compressive strength of the filler lattice must be tailored accordingly to satisfy the delocalized damage criterion. The delocalized damage criterion for a ceramic chain lattice thus follows as

$$A_r \left(\frac{L_o}{L_1}\right)^{\frac{1}{m}} \geq A_f C_1 \bar{\rho}^n, \tag{5}$$

where the left-hand side of the equation represents the tensile strength of the chain reinforcement with respect to variation in leg length $L$ (represented by $\frac{L_o}{L_1}$), and the right-hand side represents the compressive strength of the filler lattice, with $\sigma_c$ representing compressive strength of the monolithic material.

Using these constraints for delocalized damage, we can predict the upper limit of chain lattice specific energy absorption, $W_s$, starting from the general equation

$$W_s = \frac{\sigma \varepsilon}{\rho}, \tag{6}$$

where $\sigma$ is the compressive strength of the filler lattice, which is designed to be less than or equal to the tensile strength of the chain reinforcement based on the delocalized damage criteria, $\varepsilon$ is strain to failure, and $\rho$ is the overall density of the chain lattice. Assuming that the tensile and compressive strengths of a sufficiently small representative volume are equal, the maximum relative density of a filler lattice that satisfies the delocalized damage criterion can be determined by rearranging **Equation 5** to the form

$$\bar{\rho} = \left[\frac{1}{C_1}\frac{A_r}{A_f}\left(\frac{L_o}{L_1}\right)^{\frac{1}{m}}\right]^{\frac{1}{n}}. \tag{7}$$

Substituting design-dependent parameters into **Equation 6** and considering **Equation 3** gives

$$W_s = \frac{\sigma_T}{\rho_o} \cdot A_{r,T} \left(\frac{L_o}{L_1}\right)^{\frac{1}{m}} \cdot L_1 \left(1 - 1.4\left[\frac{1}{C_1}\frac{A_r}{A_f}\left(\frac{L_o}{L_1}\right)^{\frac{1}{m}}\right]^{\frac{1}{n}}\right) \cdot \frac{1}{V}, \tag{8}$$

where $\rho_o$ is the density of the monolithic material, $A_{r,T}$ is the minimum total chain reinforcement area in all planes perpendicular to the tensile loading axis, and $V$ is the total volume of material in the chain lattice unit cell.

Specific energy absorption calculations based on **Equation 8** are shown in **Figure 3**, which gives values for both bend- and stretch-dominated filler lattices. These calculations use SiOC in particular because it can be printed as lightweight, energy-absorbing lattices via stereolithography and pyrolysis of a preceramic polymer [17]. Geometric parameters are derived from the flat racetrack chain link geometry shown in **Figure 1**, following delocalized damage constraints with $\bar{\rho} = 0.3$ which is a maximum relative density for the previously described lattice strength models. There are several limitations to this prediction. First, $C_1$ is assumed as unity for the stretch-dominated chain lattice predictions, which is likely an overestimate based on compression studies of stretch-dominated lattices [18]. Second, this model assumes that all of the tensile stress in the chain reinforcement is due to compression of the filler lattice without considering contributions from other sources, such as interfacial shear, which vary based on chain lattice geometry [7]. Finally, this model assumes that the filler lattice crush-up strength is constant and equal to its compressive yield strength. Compression studies suggest that the filler lattice crush-up strength may be between roughly 40-90% of the compressive yield strength with variation during compression [19]. Nevertheless, there are several important trends we observe in **Figure 3**. Notably, each data set has a peak in specific energy absorption which demonstrates the tradeoff between increasing strain to failure and decreasing tensile strength as $L_1$ increases. Furthermore, specific energy absorption increases with Weibull modulus, with a shallower post-peak drop-off in specific energy absorption, reflecting the volume-effects of Weibull statistics on tensile strength.

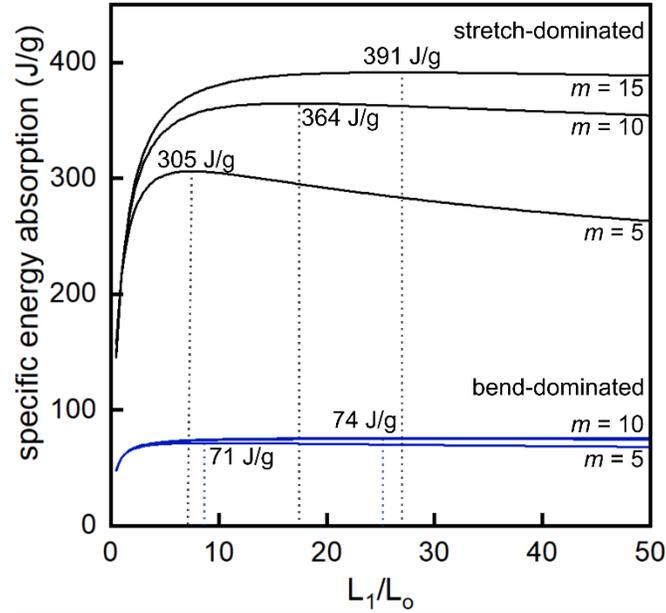

**Figure 3:** Specific energy absorption of SiOC chain lattices with different Weibull moduli and filler lattice topologies as a function of normalized chain link length $L_1/L_0$ ($\sigma_0$ = 2.6 GPa, $L_0$ = 19.21 mm, $\rho_0$ = 1.75 g/cm$^3$).

The optimal specific energy absorption values for SiOC chain lattices with relative densities ranging from 35 to 70% are shown in the property diagram in **Figure 4**, which compares the specific strength and specific energy absorption of chain lattices to reported values for ductile materials and composites as well as fully dense, 3D-printed SiOC [17]. The labels on the chain lattice datapoints indicate overall relative density. The chain lattice calculations used 3D-printed SiOC's experimentally determined Weibull modulus $m$ of 10 [20]. The calculations show that the specific energy absorption of the chain lattices increases with relative density to a maximum value (75 J/g for a bend-dominated filler lattice; 386 J/g for a stretch-dominated lattice) before starting to decrease with further increases in relative density, demonstrating there is an optimal chain lattice relative density that maximizes specific energy absorption. The optimal relative densities for chain lattices with bend- and stretch-dominated filler lattices are 0.57 and 0.55, respectively.

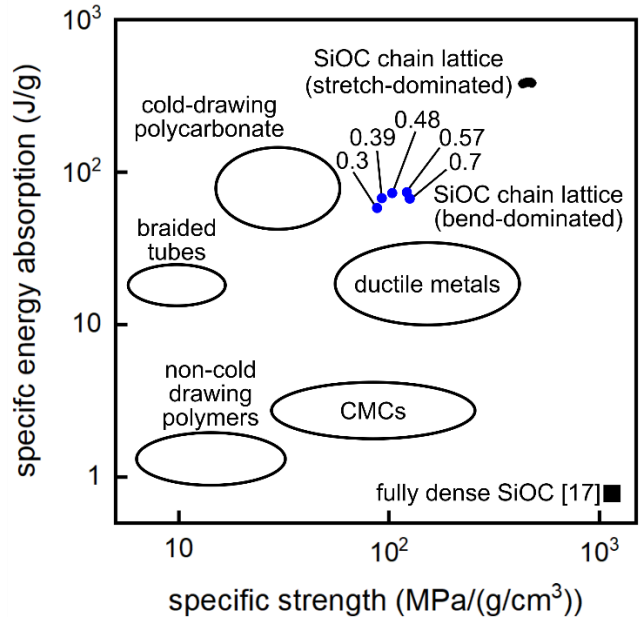

**Figure 4:** Specific energy absorption of chain lattices with bend- (blue) and stretch-dominated (black) filler lattices. The individual datapoints correspond to different relative density values. Literature data for other classes of material [7] and for fully dense 3D-printed SiOC [17] are included for comparison.

The orders-of-magnitude increment in specific energy absorption of SiOC chain lattices over that of fully dense 3D printed SiOC seen in **Figure 4** suggests chain lattices are ideal for applications where tensile energy absorption and damage-tolerance are critical. It also highlights the counterintuitive result that introducing porosity can improve energy absorption, an effect also recently observed in an elastomer [21] and in aluminum [22]. While the present work focuses on 1-D chain lattices, the concept can be extended to higher dimensions for multi-axial damage tolerance by incorporating 2-D or 3-D chainmail reinforcements. Additionally, the energy absorption characteristics can be manipulated by using other filler lattice structures, such as gyroidal triply periodic minimal surfaces [23], shell lattices [24], and honeycombs [25]. Finally, by leveraging the control over form offered by AM, chain lattices can be directly printed into complex net-shapes and incorporate functional gradients in lattice architecture that impart optimal, site-specific properties. These different capabilities, considered together, indicate that chain lattices are a promising new hierarchical lattice architecture for use with brittle, 3D-printable structural materials.

**Acknowledgements:** SVT and ZCC gratefully acknowledge support from AFRL (Award No.: FA8651-20-1-0003). ARM acknowledges support from Bechtel Oil, Gas, & Chemicals.

**Figure Captions**

**Figure 1:** Chain lattice comprising two interpenetrating lattices of the same base material. Key design features (thickness *T*, length *L*, crown radius *R*) are labeled in the magnified view of the half-unit cell.

**Figure 2:** (a) Logarithmic tensile stress-strain data for 3D-printed poly(methyl methacrylate) in different forms – fully dense, a 12% dense BCC lattice, and a chain lattice with a 12% dense BCC filler lattice. (b) Chain lattice tensile specimen with gauge length = 5.2 cm and gauge diameter = 1.3 cm. (c) Half chain link of the chain lattice, where the dimensions were as follows: *R* = 5 mm, *T* = 5 mm, and *L* = 15 mm. (d) BCC filler lattice unit cell with relative density = 12%, unit cell length = 2.5 mm, and strut width = 0.35 mm.

**Figure 3:** Specific energy absorption of SiOC chain lattices with different Weibull moduli and filler lattice topologies as a function of normalized chain link length $L_1/L_0$ ($\sigma_0$ = 2.6 GPa, $L_0$ = 19.21 mm, $\rho_0$ = 1.75 g/cm$^3$).

**Figure 4:** Specific energy absorption of chain lattices with bend- (blue) and stretch-dominated (black) filler lattices. The individual datapoints correspond to different relative density values. Literature data for other classes of material [7] and for fully dense 3D-printed SiOC [17] are included for comparison.